\documentstyle[epsfig]{aipproc}

\begin{document}
\begin{flushright}
DOE/ER/41014-22-N97
\end{flushright}

\title{Chiral Perturbation Theory in Few-Nucleon Systems
\thanks{Invited talk at the 6th Conference on the Intersections of Particle
and Nuclear Physics, Big Sky, Montana, May 1997}}

\author{U. van Kolck\thanks{Address after Jan 1 1998: 
W.K. Kellogg Radiation Laboratory, Caltech, Pasadena, CA 91125}}
\address{Department of Physics,\\ 
         University of Washington,\\ 
         Seattle, WA 98195-1560}

\maketitle

\begin{abstract}
The low-energy effective theory of nuclear physics based on chiral symmetry
is reviewed. Topics discussed include the nucleon-nucleon force, 
few-body potentials, isospin violation, pion-deuteron scattering, 
proton-neutron radiative capture, pion photoproduction on the deuteron,
and pion production in proton-proton collisions.
\end{abstract}

\section*{Introduction}
 
Although studies of its perturbative regime show that 
QCD is the theory of strong interactions,
most of the structures of nuclear physics are apparent only at low energies 
where the QCD coupling constant is not small. 
Another expansion parameter is necessary in this energy regime, and
an obvious candidate is energy itself. 

The idea of a low-energy expansion is as old as nuclear physics itself. 
Already in the 30's, Bethe and Peierls \cite{vkolck:bethe}
considered the two-nucleon system in this light. 
They based their approach on a previous argument about saturation 
due to Wigner, that the nuclear potential is of order 100 MeV and 
thus much larger than the deuteron binding energy of 2.2 MeV, but
has a range $R$ ($\simeq$ 1.4 fm) much smaller than the size 
$\lambda$ of the deuteron ($\simeq$ 4.4 fm).
Bethe and Peierls reasoned that as a consequence, for distances $r$ such that 
$R\lesssim r \lesssim \lambda$, only s-waves are important and the sole 
effect of the potential is to provide an energy independent 
boundary condition at $r\sim R$.
Up to an error of $O(R/\lambda)$, then, the system could be described by 
a free Schr\"odinger equation with the boundary condition that 
the logarithmic derivative of the radial wavefunction is a constant at $r=0$. 
This constant cannot be calculated without detailed knowledge of 
the potential, so it was fitted to the deuteron binding energy. 
If that was all, nothing would have been learned.
Bethe and Peierls' point, however, was that they could then predict (with a 
30\% uncertainty) other processes 
---such as $pn\rightarrow pn$, $\gamma d\rightarrow pn$,
$\gamma d\rightarrow \gamma d$, and $ed\rightarrow e'pn$--- 
at energies comparable to the deuteron binding energy.

This approach can be rephrased in an effective field theory (EFT) language.
At typical momenta $Q$ much smaller than the pion mass $m_\pi$, 
the relevant degree of freedom is the nucleon,
the important symmetries are parity, time-reversal and Galilean symmetry, 
and the appropriate expansion parameter is $Q/m_\pi \sim R/\lambda$.
(Electromagnetic processes can also be considered by adding the photon,
$U(1)_{em}$ gauge invariance, and $\alpha_{em}$ to this list.) 
The most general Lagrangian involving nucleons only consists of 
an infinite number of terms, which are quadratic, quartic, ...,  
in the nucleon fields with increasing number of derivatives. 
By dimensional reasons, derivatives come associated 
with inverse factors of a mass scale $m_\pi$ or greater.
Nucleons are non-relativistic and the corresponding field theory has 
nucleon number conservation. 
The $T$-matrix for the two-nucleon system is simply a sum of bubble graphs, 
whose vertices are the four-nucleon contact terms that appear 
in the Lagrangian. Formally, this is equivalent to solving a Schr\"odinger 
equation with a low-energy, effective potential
consisting, schematically, of a sum 
$\bar{C}_0 \delta(\vec{r}) + \bar{C}_2 \delta''(\vec{r}) +...$, 
where the $\bar{C}_n$'s are the coefficients of the contact terms,
expected to scale as $\bar{C}_n \sim \bar{C}_0/ m_\pi^n$.
The net effect is thus to replace the ``true'', possibly complicated potential
of range $\sim 1/m_\pi$ by a multipole expansion with moments $\bar{C}_n$. 
Life is somewhat more complicated, however, because this field theory, 
like any other, has to be renormalized. The bubbles are actually ultraviolet 
divergent, requiring regularization and absorption of the regulator 
dependence in renormalized parameters. 
It is not difficult to show \cite{vkolck:vkolck1} that the
effect of renormalization is to turn the effective potential into a 
generalized pseudo-potential, or equivalently, turn the problem into a free 
one with boundary conditions at the origin which are analytic in the energy. 
The first, energy-independent term, 
parametrized by $\bar{C}_0^R$, is just the one considered by Bethe and Peierls.
Much effort has been spent during the last year in trying to understand 
issues related to regularization and fine-tuning of 
this ``pionless'' theory \cite{vkolck:kaplan}.

To the extent that there are no assumptions about the detailed dynamics of 
the ``underlying'' theory, the effective theory cannot be wrong and it is 
useful as long as $Q\ll m_\pi$. One may ask how strong this restriction is,
however, when we consider physics of more than two nucleons. 
Clearly, we can always apply the pionless theory 
to sufficiently low-energy scattering situations, 
where we can control the momenta of the initial and final nucleons. 
But one noble goal of nuclear physics is to understand nuclei themselves. 
Let me use $2 m_N B/ A$ as a measure of a typical momentum $Q$ of 
a nucleon of mass $m_N$ in a nucleus with $A$ nucleons and binding energy $B$. 
(Other quantities such as charge radii give similar estimates.) 
$Q/m_\pi$ is then about 0.3 for $^2$H, 0.5 for $^3$H, 0.8 for $^4$He, ..., 
and 1.2 for symmetric nuclear matter in equilibrium. 
The same argument that justified the use of a pionless theory for the deuteron
now suggests that understanding the binding of typical nuclei ($^4$He 
and heavier) requires {\it explicit} inclusion of pions, 
but {\it not} of heavier mesons such as the rho. 

Now, we are in luck because QCD does explain the special role of the pion, and 
in the process, provides a rationale to treat pion effects systematically. 
This procedure goes by the name of Chiral Perturbation Theory ($\chi$PT); 
it will be described briefly in the next section, and 
exemplified even more briefly in the simplest case of at most one nucleon 
in the following section. I then come to the main portion of this review, 
where we tackle nuclear forces and external probes of light nuclei.

\section*{Effective Chiral Lagrangian}

Why is the pion special? It is one of the nicest features of QCD that it
provides a scenario where the lightness of the pion results 
from its (pseudo-) Goldstone boson nature.
Here for simplicity I will limit myself to the case of two quark flavors.

If the quark masses were zero (``chiral limit''), 
the QCD Lagrangian would be invariant under transformations of the group 
$SU(2)_L\times SU(2)_R$ of independent rotations of the quarks' 
left- and right-handed components. 
When acting on quark bilinears, this chiral symmetry is equivalent to $SO(4)$. 
A quick look at the hadronic spectrum convinces us that the chiral limit 
can only be phenomenologically relevant if there is spontaneous breakdown 
of chiral symmetry down to its diagonal subgroup, 
the $SU(2)_{L+R}$ ($\sim SO(3)$ for bilinears) of isospin.
Although the mechanism of spontaneous breaking is not 
sufficiently understood at the present for a detailed derivation, 
we know that the effective QCD potential as a function of four quark 
bilinears $(\bar{q}\gamma_5\tau_i q, \bar{q}q)$ has to have roughly
a mexican hat shape, i.e. $SO(4)$ symmetry with minima in a
``chiral circle'' away from the origin.

\newcommand{\boldpi}{\mbox{\boldmath $\pi$}}
Goldstone's theorem assures us there is in the spectrum, as a consequence, 
a (pseudo-)scalar boson, which corresponds to excitations 
in the coset space $SO(4)/SO(3) \sim S^3$. 
We call the radius of this ``circle'' $f_\pi$,
which is a function of $\Lambda_{QCD}$ 
that ends up being the pion decay constant $\simeq 92$ MeV.
At sufficiently low energies it is convenient to assign a field $\boldpi$
to the pion in the effective Lagrangian. 
An infinitesimal chiral transformation is of the form 
$\boldpi\rightarrow\boldpi + f_\pi \mbox{\boldmath $\epsilon$}+...$ .
$SO(4)$ symmetry of the dynamics implies that the Lagrangian will have a 
piece that is a function of $\boldpi$ only through derivatives of $\boldpi$
on the circle, which are non-linear. The Lagrangian, in principle
completely determined by $\Lambda_{QCD}$, has an infinite number of terms
with arbitrarily high pion self-interactions, but without a pion mass term. 

We know, however, that not all quark masses are zero. 
Quark masses generate two terms in the QCD Lagrangian. 
One term, $\bar{m} \bar{q}q$ with $\bar{m}=(m_u+m_d)/2$, is the
fourth component of an $SO(4)$ vector and therefore breaks $SO(4)$ 
explicitly down to $SO(3)$ of isospin. It causes a tilt of 
the effective potential in the $\bar{q}q$ direction determined
by the small parameter $\eta=\bar{m}/\Lambda_{QCD}$.
The effective Lagrangian will acquire then a piece 
that breaks $SO(4)$ explicitly in the same way as $\bar{q}q$. 
This piece is another infinite set of terms
that do depend on $\boldpi$ in an isospin invariant way, 
but are all proportional to powers of $\eta$. 
In particular, $m_\pi^2\propto \eta \Lambda_{QCD}^2$.
The other quark mass term, $\epsilon \bar{m} \bar{q}\tau_3 q$ with 
$\epsilon=(m_u-m_d)/(m_u+m_d)\simeq 1/3$, is the third component of 
another $SO(4)$ vector and further breaks $SO(3)$ down to $U(1)\times U(1)$. 
Likewise, the effective Lagrangian will inherit yet another 
infinite set of terms, this time that break isospin as $\bar{q}\tau_3 q$
and are, in principle, of order $\epsilon$ relative 
to the isospin conserving chiral breaking effects. Why isospin breaking is 
in fact much smaller in most observables will be explained later.

QCD therefore has all the ingredients to provide a rationale not only for 
the special role of the pion, but also for a systematic treatment of its 
effects: pion interactions are weak at low energies 
due to (approximate) chiral symmetry.

We can now formulate an EFT for momenta 
$Q\sim m_\pi\ll M_{QCD}\sim m_\rho \sim m_N \sim 4\pi f_\pi$
along the same lines of the pionless case. 
The extra degrees of freedom 
---besides non-relativistic nucleons and photons---
are obviously pions and also non-relativistic delta isobars, 
since the delta-nucleon mass difference $m_{\Delta}-m_N\sim 2m_\pi$ 
is of the order of the momenta $Q$ we want to consider. 
The new and very important symmetry is approximate $SU(2)_L\times SU(2)_R$. 
The expansion parameter
is expected to be $Q/M_{QCD}$ ---besides $\alpha_{em}$. 
The most general Lagrangian with these ingredients has schematically the form 
\begin{eqnarray}
{\cal L} & = & \sum_{\{dqnpf\}=1}^{\infty} C_{dqnpf}
               (\frac{{\cal D}}{M_{QCD}})^d 
               (\frac{m_\Delta -m_N}{M_{QCD}})^q
               (\frac{m_\pi}{M_{QCD}})^n
               (\frac{\boldpi}{f_\pi})^p
               (\frac{\psi^\dagger \psi}{f_\pi^2 M_{QCD}})^{\frac{f}{2}}
               f_\pi^2 M_{QCD}^2 \nonumber \\
 & = & \sum_{\Delta=0}^{\infty} {\cal L}_{(\Delta)}.  \label{E:vkolck:L}
\end{eqnarray}
Here the $C$'s are parameters assumed to be natural, i.e.
of $O(1)$ ---or $O(\epsilon^\#)$ with $\#$ a positive integer, in the case 
of isospin breaking operators originating in the quark mass difference. 
$\psi$ stands for a nucleon or delta isobar, 
and $\cal D$ for a covariant derivative.
The interactions are naturally grouped in sets ${\cal L}_{(\Delta)}$ 
of common index $\Delta\equiv d+q+n+\frac{f}{2} -2$ 
but arbitrary values of $p$. 
For non-electromagnetic interactions, we find that 
$\Delta\ge 0$ only because of chiral symmetry. 

Consider an arbitrary irreducible contribution to a process involving
$A$ nucleons and an arbitrary number of pions and photons, all with momenta 
of order $Q$. 
It can be represented by a Feynman diagram with $A$ continuous nucleon lines,
$L$ loops, $C$ separately connected pieces, and 
$V_\Delta$ vertices from ${\cal L}_{(\Delta)}$, 
whose connected pieces cannot be all split by cutting only nucleon lines.
($C=1$ for $A=0,1$; $C=1, ..., A-1$ for $A\ge 2$. The reason to consider
irreducible diagrams and $C> 1$ will be discussed in the section
on few-nucleon systems.) 
It is easy to show \cite{vkolck:weinberg1} that this
contribution is typically of $O((Q/M_{QCD})^\nu)$, where
\begin{equation}
\nu=4-A+2(L-C)+\sum_\Delta V_\Delta \Delta.  \label{E:vkolck:nu}
\end{equation}
Since $L$ is bounded from below (0) and $C$ from above ($C_{max}$), 
the chiral symmetry constraint $\Delta\ge 0$
implies that $\nu\ge \nu_{min}=4-A-2C_{max}$ for strong interactions. 
Leading contributions come from tree diagrams built out of ${\cal L}_{(0)}$
and coincide with current algebra. Perturbation theory in $Q/M_{QCD}$
can be carried out by considering contributions from ever increasing $\nu$.

Note that this approach is:

({\it i}) systematic. It is a {\it perturbation} in the number of loops, 
derivatives/fermion fields, and many-nucleon effects.

({\it ii}) consistent with QCD. The only (very important) QCD inputs are
confinement (color singlet fields), symmetries ({\it chiral}, ...), and
naturalness. QCD can be represented by a point in the 
space of renormalized parameters (at some renormalization scale) of
the effective theory. 
An explicit solution of QCD (such as from simulations on a lattice)
would provide knowledge of the exact position of this point, 
and then the effective theory would be completely predictive. 
Until such a solution is found ---or as a test of QCD after it is found--- 
we can recourse to fitting low-energy experiments in order to determine 
the region in parameter space allowed on phenomenological grounds. 
Even in this case the theory is predictive, because
to any given order the space of parameters is finite-dimensional.
After a finite number of experimental results are used, an infinite number 
of others can be predicted up to an accuracy depending on the order of the 
expansion. In practice, because the number of parameters grows rapidly 
with the order, model-dependent estimates of parameters based on 
specific dynamic ideas 
---such as saturation by tree-level resonance exchange--- are sometimes used.

({\it iii}) a {\it theory}. It is applicable in principle to all low-energy 
phenomena. I will in the next section mention some of the highlights of 
this program for processes with at most one nucleon, and then most of the 
rest of the paper is devoted to nuclear physics {\it per se}.

\section*{Meson and One-nucleon Sectors}

In the case of strong mesonic processes, 
$\nu=2+2L+\sum_\Delta V_\Delta \Delta\geq 2$
with $\Delta= d+n-2$ increasing in steps of two. 
Since Weinberg discovered this systematic generalization of current algebra
\cite{vkolck:weinberg2} 
and Gasser and Leutwyler implemented it
\cite{vkolck:gasser1},
many processes (including weak interactions)
involving pions (and kaons) have been examined, typically to $\nu=4$. 
The most thoroughly studied process has been $\pi\pi$ scattering,
where a $\nu=6$ calculation was carried out \cite{vkolck:bijnens};
fits to other data plus resonance saturation show convergence and
good agreement with phase shifts through energies 
of more than 100 MeV above threshold.
Many good reviews exist
on the mesonic sector; see for example Ref. \cite{vkolck:meson}.

For processes where one nucleon is present,    
$\nu=1+2L+\sum_\Delta V_\Delta \Delta \ge 1$, where
$\Delta= d+q+n+\frac{f}{2} -2$ (with $f=0, 2$) increases in steps of one.
(Convergence can be expected to be slower compared to purely mesonic 
interactions.)
Thanks to this power counting, a low-energy nucleon can in a very 
definite sense be pictured as a static, point-like object 
(up to corrections in powers of $Q/m_N$), surrounded by: 
i) an inner cloud which is dense but of short range $\sim 1/m_\rho$, 
so that we can expand in its relative size $Q/m_\rho$;
ii) an outer cloud of long range $\sim 1/m_\pi$ but sparse,
so that we can expand in its relative strength $Q/4 \pi f_\pi$. 

The first attempt at including the nucleon in $\chi$PT
\cite{vkolck:gasser2} had limited success because it did not fully explore
the non-relativistic nature of the nucleon and did not consider
the delta isobar explicitly. This was remedied in the work of
Jenkins and Manohar \cite{vkolck:jenkins}, while a more complete 
analysis of the 
role of the delta in one-nucleon process has been carried out recently
\cite{vkolck:hemmert}. 
However, most calculations have been limited to the threshold region
where the delta contribution is relatively unimportant and a
``deltaless'' theory useful; see for example
Ref. \cite{vkolck:nucleon} for an extensive review.

Here, for illustration, I mention explicitly the case of pion 
photoproduction at threshold which has received a lot of attention lately.
At threshold the amplitude for a photon of 
polarization $\vec{\epsilon}$ incident on a nucleon of spin $\vec{\sigma}$ is 
$T\sim i \vec{\sigma}\cdot \vec{\epsilon} E_{0+}$. 
This process has been studied up to $\nu=4$ in the deltaless theory
in Ref. \cite{vkolck:bernard1} (where references to the experimental 
papers can be found; see also Ref. \cite{vkolck:korkmaz}). 
Results for $E_{0+}$ from a fit constrained by resonance saturation 
are presented in Table \ref{T:vkolck:E0+}.

\begin{table}[b!]
\caption{Values for $E_{0+}$ in units of $10^{-3}/m_{\pi^+}$
at various orders $\nu$ in $\chi$PT and in experiments, for the 
different channels.}
\label{T:vkolck:E0+}
\begin{tabular}{lccccc}
$E_{0+}$($10^{-3}/m_{\pi^+}$)& $\nu=1$ & $\nu=2$ & $\nu=3$ & $\nu=4$ 
                                                         & Experiment\\
\tableline
$\gamma p\rightarrow \pi^+ n$ &    34.0 &    26.4  &    28.9  &    28.2
                                                   &    27.9 $\pm$ 0.5.\\
                              &         &          &          &        
                                                   &    28.8 $\pm$ 0.7.\\
$\gamma n\rightarrow \pi^- p$ & $-$34.0 & $-$31.5  & $-$32.9  & $-$32.7 
                                                   & $-$31.4 $\pm$ 1.3. \\
                              &         &          &          &
                                                   & $-$32.2 $\pm$ 1.2. \\
$\gamma p\rightarrow \pi^0 p$ &     0   &  $-$3.58 &     0.96 &  $-$1.16
                                                   &  $-$1.31 $\pm$ 0.08. \\
                              &         &          &          &
                                                   &  $-$1.32 $\pm$ 0.08. \\
$\gamma n\rightarrow \pi^0 n$ &     0   &     0    &     3.7  &    2.13
                                                   &            ? \\
\end{tabular}
\end{table}

We can observe that the big size of the charged pion channels result
from a non-vanishing $\nu=1$ contribution (the Kroll-Ruderman term). 
Convergence and agreement with experimental values are pretty good.
For the neutral pion channels convergence is less apparent, but
the absolute values much smaller. 
The $\nu=4$ result for $\gamma p\rightarrow \pi^0 p$ is in 
relatively good agreement with the recent results from Mainz and Saskatoon. 
To the same order there is a prediction for the 
$\gamma n\rightarrow \pi^0 n$ reaction, which would be important to test. 
If isospin symmetry breaking is neglected, there are
only three independent amplitudes; if we use the three measured amplitudes,
their uncertainties limit extraction of $E_{0+}(\pi^0 n)$ 
to the range $-0.5\mapsto +2.5$. 
Newer, more accurate data for the charged channels were presented
at this conference by Korkmaz \cite{vkolck:korkmaz}.
Isospin breaking entered the calculation
of Table 1, albeit in an incomplete form. 
A direct measurement of $E_{0+}(\pi^0 n)$ could not only
check the consistency of the calculation, but also provide a new
isospin violating observable. This however requires a deuteron target,
and a description of $\gamma d\rightarrow \pi^0 d$ in the same framework.
So, even from the point of view of nucleon properties 
we are led naturally into the study of light nuclei.

\section*{Nuclear Physics}

A non-trivial new element enters the theory when we consider systems of
more than one nucleon \cite{vkolck:weinberg1}. 
Because nucleons are heavy, contributions from intermediate states
that differ from the initial state only in the energy of nucleons are
enhanced by infrared quasi-divergences. These are linked to the existence of
small energy denominators of $O(Q^2/m_N)$, which generate contributions
$O(m_N/Q)$ larger than what would be expected from Eq. (\ref{E:vkolck:nu}).
The latter is still correct for the class of sub-diagrams
---called irreducible--- that do not contain intermediate states with
small energy denominators. 
For an $A$-nucleon system these are $A$-nucleon irreducible diagrams, 
the sum of which we call the potential $V$.
When we consider external probes all with $Q\sim m_\pi$, 
the sum of irreducible diagrams forms the kernel $K$ 
to which all external particles are attached.
A generic diagram contributing to a full amplitude will consist
of irreducible diagrams sewed together by states of small energy denominators. 
These irreducible diagrams might have more than one connected piece, 
hence the introduction of $C$ in Eq. (\ref{E:vkolck:nu}).
The infrared enhancement requires that we sum diagrams to all orders
in the amplitude, creating the possibility of the existence 
of shallow bound states (nuclei). For an $A$-nucleon system, 
this is equivalent to solving the 
Schr\"odinger equation with the potential $V$. The amplitude for a process
with external probes is then $T\sim \langle \psi'|K|\psi\rangle$ 
where $|\psi\rangle$ ($|\psi'\rangle$) 
is the wavefunction of the initial (final) nuclear state calculated 
with the potential $V$.

Because our $Q/M_{QCD}$ expansion is still valid 
for the potential and the kernel,
the picture of a nucleon as a mostly static object surrounded by an inner
and an outer cloud leads to remarkable nuclear physics properties that 
we are used to, but would remain otherwise not understood from the 
viewpoint of QCD.

\subsection*{Nuclear Forces}

If we put a few non-relativistic nucleons together, each nucleon will not
be able to distinguish details of the others' inner clouds. The region of 
the potential associated with distances of $O(1/m_{\rho})$ can be expanded in
delta-functions and their derivatives as Bethe and Peierls did. The outer 
cloud of range $O(1/m_{\pi})$ yields non-analytic contributions
to the potential, but being sparse, 
it mostly produces the exchange of one pion, 
with progressively smaller two-, three-, ...- pion exchange contributions. 

\begin{figure}[t]
\centerline{\epsfig{file=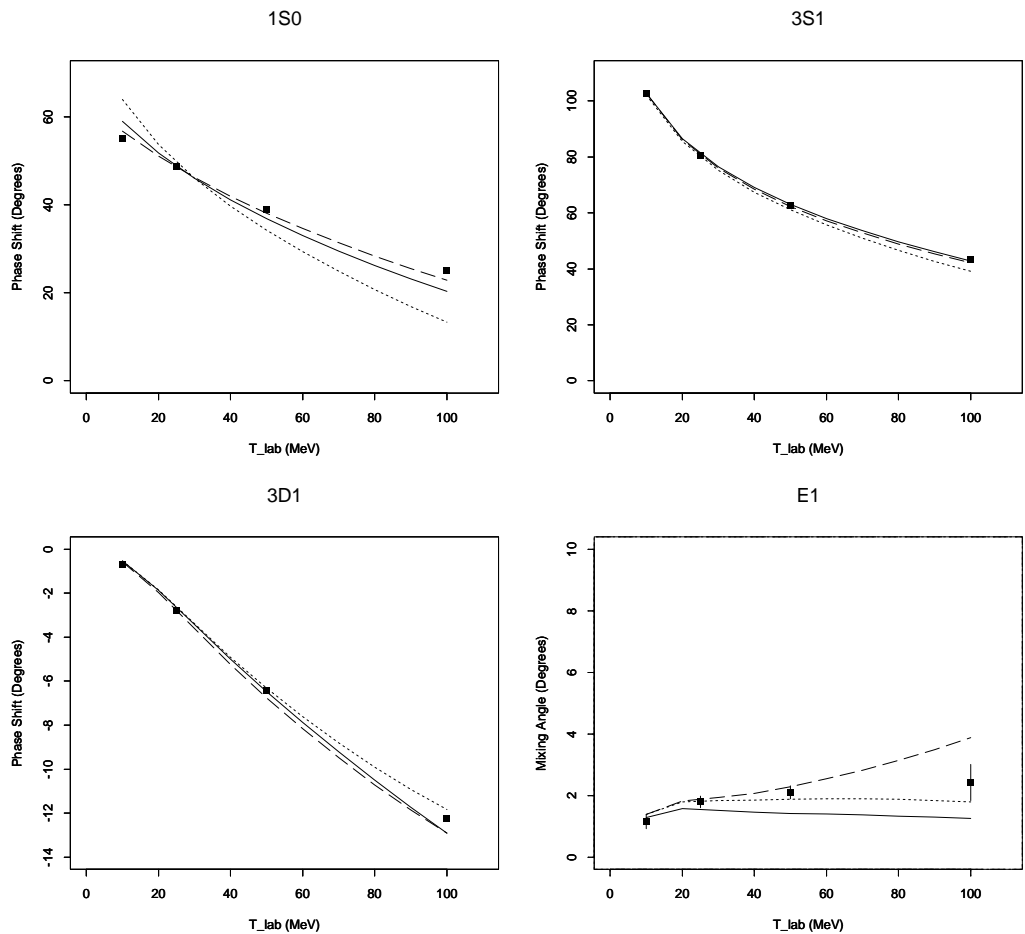,height=7.5in,width=5.5in}}
\vspace{-6cm}
\caption{$^1S_0$, $^3S_1$, and $^3D_1$ $NN$ phase shifts and 
$\epsilon_1$ mixing angle in degrees as functions
of the laboratory energy in MeV: 
chiral expansion up to $\nu=3$ for cut-offs of 
500 (dotted), 780 (dashed), and 1000 MeV (solid line); 
and Nijmegen phase shift analysis (squares).}
\label{F:vkolck:NNfigs}
\end{figure}

For the two-nucleon system, 
$\nu=2L+\sum_\Delta V_\Delta \Delta$, with $\Delta$ as in the 
one-nucleon case. 
A calculation of all the contributions up to $\nu=3$ was carried out
in Ref.\cite{vkolck:ordonez}. 
In leading order, $\nu=0$, the potential is simply static one-pion exchange 
and momentum-independent contact terms \cite{vkolck:weinberg1}. 
$\nu=1$ corrections vanish due to parity and time-reversal invariance. 
$\nu=2$ corrections include several two-pion exchange diagrams 
(including virtual delta isobar contributions), recoil in one-pion exchange, 
and several contact terms that are quadratic in momenta. 
At $\nu=3$ a few more two-pion exchange diagrams have to be considered. 
As in the pionless case, regularization and renormalization are necessary. 
It is not straightforward to implement dimensional regularization in this 
non-perturbative context, so we used an overall gaussian cut-off, 
and performed calculations with the cut-off parameter
$\Lambda$ taking values 500, 780 and 1000 MeV. 
Cut-off independence means that for each cut-off value a set of 
(bare) parameters can be found that fits low-energy data. 
A sample of the results for the lower, more important partial waves
is presented in Fig. \ref{F:vkolck:NNfigs} and for the deuteron quantities in 
Table \ref{T:vkolck:dparam}. (See \cite{vkolck:ordonez} for more details
and reference to experiments and phase shift analyses.)

\begin{table}[b!]
\caption{Effective chiral Lagrangian fits for various cut-offs $\Lambda$
and experimental values 
for the deuteron binding energy ($B$), magnetic moment ($\mu_d$), electric 
quadrupole moment ($Q_E$), asymptotic $d/s$ ratio ($\eta$), and $d$-state 
probability ($P_D$).} 
\label{T:vkolck:dparam}
\begin{tabular}{ccccc}
Deuteron   & \multicolumn{3}{c}{$\Lambda$ (MeV)} &            \\
\cline{2-4}
quantities & 500 & 780 & 1000                   & Experiment \\
\tableline
$B$ (MeV)  & 2.15 & 2.24 & 2.18                 & 2.224579(9) \\
$\mu_d$ ($\mu_N$) & 0.863 & 0.863 & 0.866       & 0.857406(1) \\
$Q_E$ (fm$^2$) & 0.246 & 0.249 & 0.237          & 0.2859(3)   \\
$\eta$     & 0.0229 & 0.0244 & 0.0230           & 0.0271(4)  \\
$P_D$ (\%) & 2.98 & 2.86 & 2.40                 &            \\
\end{tabular}
\end{table}

The fair agreement of this first calculation and data 
up to laboratory energies of 100 MeV or so suggests that this may 
become an alternative to other, more model-dependent approaches 
to the two-nucleon problem.
Further examination of regularization effects, fine-tuning in the $^1S_0$
channel, and different aspects of two-pion exchange in this context can be 
found in Refs. \cite{vkolck:kaplan,vkolck:2pion}.

Perhaps more impressive is that we can get some insight into other aspects 
of nuclear forces. 
Let us look for the new forces that appear in systems with more than two 
nucleons. 
The dominant potential, at $\nu=6-3A=\nu_{min}$, 
is the two-nucleon potential of lowest order that appeared in 
the two-nucleon case.
We can easily verify that a three-body potential will arise at 
$\nu=\nu_{min}+2$, a four-body potential at $\nu=\nu_{min}+4$, and so on. 
It is (approximate) chiral symmetry therefore that
implies that $n$-nucleon forces $V_{nN}$ are expected to obey a hierarchy
of the type
$\langle V_{(n+1)N}\rangle/\langle V_{nN}\rangle
\sim O((Q/M_{QCD})^2)$,
with $\langle V_{nN}\rangle$ denoting the contribution per $n$-plet.
If we estimate 
$\langle V_{2N}\rangle \sim \frac{g_A^2}{16\pi f_\pi^2} m_\pi^3 \simeq 10$ MeV,
we can guess 
$\langle V_{3N}\rangle \sim .5$ MeV,
$\langle V_{4N}\rangle \sim .02$ MeV, and so on. 
This is in accord with detailed few-nucleon calculations using more 
phenomenological potentials.
The explicit three-body potential at $\nu=\nu_{min}+2$ (from the delta 
isobar) and $\nu_{min}+3$ was derived in Ref.\cite{vkolck:vkolck2}.

We can also look at isospin-dependent effects. Within the context of
$\chi$PT it can be shown \cite{vkolck:vkolck3}
that isospin is an accidental 
symmetry, in the sense that it does not appear in the low-energy EFT
in lowest order, and therefore is typically not an $O(\epsilon)$ effect,
but $O(\epsilon (Q/M_{QCD})^n), n\ge 1$.
In the case of nuclear forces, we find moreover a hierarchy among 
different types of components. It is standard to call 
class I the strongest forces that are isospin symmetric, 
class II weaker forces that are isospin violating but charge symmetric, 
class III even weaker forces that are charge symmetry breaking but
symmetric under permutation of particles in isospin space, 
and class IV the weakest, remaining forces. 
In the chiral expansion, one indeed finds \cite{vkolck:vkolck3} 
that higher class forces appear at higher orders:
$\langle V_{\rm M+1}\rangle /\langle V_{\rm M}\rangle
\sim O(Q/M_{QCD})$,
where $\langle V_{\rm M}\rangle$ denotes the contribution of 
the leading class ${\rm M}$ potential. 
This qualitatively explains, for example, the observed isospin structure 
of the two-nucleon Coulomb-corrected scattering lengths, 
$a_{np}\simeq 4 \times ((a_{nn}+a_{pp})/2 - a_{np}) 
       \simeq 4^2 \times (a_{pp}-a_{nn})$. 
Precise calculations of simultaneous electromagnetic and strong 
isospin violation in the nuclear potential have also been carried out
\cite{vkolck:vkolck4}.   

Despite these successful fits and insights, 
the main advantage of the method of EFT lies 
in its concomitant application to many other processes, which
might yield more predictive statements. I now discuss some of these.

\subsection*{Nuclear Probes}

Before plunging into hard results, let me point out another generic result
of the chiral expansion. 
As a result of the factor $-2C$ in Eq. (\ref{E:vkolck:nu}), we
see immediately ---in an effect similar to few-nucleon forces--- that
external low-energy probes ($\pi$'s, $\gamma$'s) will tend to interact 
predominantly with a single nucleon, simultaneous interactions with more than 
one nucleon being suppressed by powers of $(Q/M_{QCD})^2$. Again, this is a 
well-known result that arises naturally here. 

This is of course what allows extraction, to a certain accuracy, of  
one-nucleon parameters from nuclear experiments. 
More interesting from the nuclear dynamics perspective are, however,
those processes where the leading single-nucleon contribution vanishes by 
a particular choice of experimental conditions, 
for example the threshold region. 
In this case the two-nucleon contributions, especially in the 
relatively large deuteron, can become important.

\paragraph*{$\pi d\rightarrow \pi d$ at threshold.}
This is perhaps the most direct way to check the consistency of $\chi$PT in
few-nucleon systems and in pion-nucleon scattering.
Here the lowest-order, $\nu=-2$ contributions to the kernel vanish because 
the pion is in an s-wave and the target is isoscalar.
The $\nu=-1$ term comes from the (small) isoscalar pion-nucleon seagull,
related in lowest-order to the pion-nucleon isoscalar amplitude $b_0$.
$\nu=0, +1$ contributions come from corrections to pion-nucleon scattering
and two-nucleon diagrams, which involve besides $b_0$ also the much larger 
isovector amplitude $b_1$. Weinberg\cite{vkolck:weinberg3} has estimated
these various contributions to the pion-deuteron scattering length, finding 
agreement with previous, more phenomenological calculations,
which have been used to extract $b_0$.

\paragraph*{$n p \rightarrow \gamma d$ at threshold.}
This offers a chance of a precise postdiction.
Here it is the transverse nature of the real outgoing photon that is 
responsible for the vanishing of the lowest-order, $\nu=-2$ contribution
to the kernel. 
The single-nucleon magnetic contributions come 
at $\nu=-1$ (tree level), $\nu=+1$ (one loop), etc. 
The first two-nucleon term is an one-pion exchange at $\nu=0$
long discovered to give a smaller but non-negligible contribution.
There has been a longstanding discrepancy of a few percent between these
contributions and experiment. 
At $\nu=+2$ there are further one-pion exchange, two-pion exchange,
and short-range terms. 
Park, Min and Rho \cite{vkolck:park1} calculated the two-pion exchange
diagrams in the deltaless theory and used resonance saturation to estimate 
the other $\nu=+2$ terms. With wavefunctions from the Argonne V18 
potential and a cut-off $\Lambda=1000$ MeV,
they found the excellent agreement with experiment shown in 
Table \ref{T:vkolck:radcap}. 
The total cross-section changes by .3 \% 
if the cut-off is decreased to 500 MeV. 
(See \cite{vkolck:park1} where reference to experiment can be found.)

\begin{table}[b!]
\caption{Values for various contributions to 
the total cross-section $\sigma$ for radiative neutron-proton capture in mb:
impulse approximation to $\nu=2$ ($^{imp}$),
impulse plus two-nucleon diagrams at $\nu=0$ ($^{imp+tn0}$),
impulse plus two-nucleon diagrams up to $\nu=2$ ($^{imp+tn}$),
and experiment ($^{expt}$).} 
\label{T:vkolck:radcap}
\begin{tabular}{l|d|d|d}
$\sigma^{imp}$ & $\sigma^{imp+tn0}$ & $\sigma^{imp+tn}$ & $\sigma^{expt}$\\
\tableline
305.6 &  321.7 & 336.0  &  334$\pm$0.5  \\
\end{tabular}
\end{table}

\paragraph*{$\gamma d\rightarrow \pi^0 d$ at threshold.}
As emphasized earlier, this reaction offers the possibility to test a
prediction arising from a combination of two-nucleon contributions 
and the neutral pion single-neutron amplitude. 
Here, it is the neutrality of the outgoing s-wave pion that ensures
that the leading $\nu=-2$ terms vanish. 
The single-scattering contribution is given by the same $\nu=-1, 0, +1,...$
mechanisms described earlier, with due account of p-waves and 
Fermi motion inside the deuteron. 
The first two-nucleon term enters at $\nu=0$, a correction 
appears at $\nu=+1$, and so on.  
At threshold the amplitude for a photon of polarization $\vec{\epsilon}$ 
incident on the deuteron of spin $\vec{J}$ is 
$T\sim 2i \vec{J}\cdot \vec{\epsilon} E_{d}$.
Results for $E_{d}$ up to $\nu=+1$ have been obtained \cite{vkolck:beane}
and are shown in Table \ref{T:vkolck:Ed}.
They correspond to the Argonne V18 potential and a cut-off 
$\Lambda=1000$ MeV. Other realistic potentials and cut-offs from 650 to
1500 MeV give the same result within 5\%, 
while a model-dependent estimate \cite{vkolck:wilhelm} of some $\nu=+2$ terms 
suggests a 10\% or larger error from the neglected higher orders 
in the kernel itself. 
The single-scattering amplitude depends on $E_{0+}(\pi^0 n)$
in such a way that $E_{d} \sim -1.79 -0.38(2.13-E_{0+}(\pi^0 n))$ 
in units of $10^{-3}/m_{\pi^+}$.
Thus, some sensitivity to $E_{0+}(\pi^0 n)$ survives the large two-nucleon
contribution at $\nu=0$. 

\begin{table}[t]
\caption{Values for $E_{d}$ in units of $10^{-3}/m_{\pi^+}$ from 
single scattering up to $\nu=1$ ($^{ss}$),
two-nucleon diagrams at $\nu=0$ ($^{tn0}$), 
two-nucleon diagrams at $\nu=1$ ($^{tn1}$),
and their sum ($^{ss+tn}$).}
\label{T:vkolck:Ed}
\begin{tabular}{ldd|d}
$E_{d}^{ss}$ & $E_{d}^{tn0}$ & $E_{d}^{tn1}$ & $E_{d}^{ss+tn}$ \\
\tableline
0.36 &  $-$1.90 & $-$0.25  &  $-$1.79  \\
\end{tabular}
\end{table}

Some old Saclay data gave $E_{d}=-(1.7\pm 0.2)\cdot 10^{-3}/m_{\pi^+}$ 
on its latest reanalysis, but a new precise measurement is called for.
A test of the above prediction will come from new Saskatoon data, 
currently under analysis \cite{vkolck:korkmaz}.
An electroproduction experiment will also be carried out in Mainz
\cite{vkolck:bernstein}.

\begin{figure}[t] 
\centerline{\epsfig{file=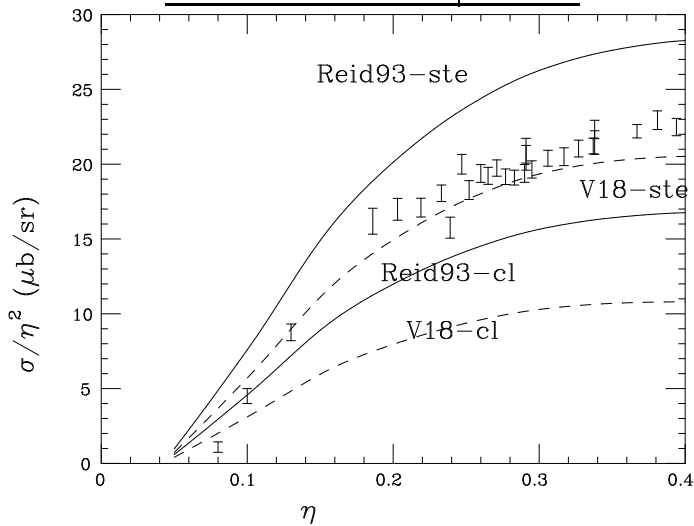,angle=270,height=2.5in,width=2.5in,
                    bbllx=1.5in,bblly=2.5in,bburx=7.5in,bbury=8.5in}}
\vspace{.5cm}
\caption{Cross-section for $pp \rightarrow pp \pi^0$ in $\mu$b/sr
as function of the pion momentum $\eta$ in units of $m_\pi$ for 
two $NN$ potentials (Argonne V18 and Reid93) 
and two parametrizations of the isoscalar pion-nucleon amplitude (ste and cl).}
\label{F:vkolck:mesexfig3}
\end{figure}

\paragraph*{$pp \rightarrow pp\pi^0$ close to threshold.}
This reaction has attracted a lot of interest because of the failure of 
standard phenomenological mechanisms in reproducing 
the small cross-section near threshold. 
It involves larger momenta of $O(\sqrt{m_\pi m_N})$, so the 
the relevant small parameter here is the not so small
$(m_\pi/m_N)^{\frac{1}{2}}$.
It is therefore not a good testing ground for the above ideas. 
But $(m_\pi/m_N)^{\frac{1}{2}}$ is still $<1$, so at least in some formal sense
we can perform a low-energy expansion. 
In Ref.\cite{vkolck:cohen}
the chiral expansion was adapted to this reaction and the first few 
contributions estimated. 
Again, the lowest order terms all vanish. The formally leading non-vanishing 
terms ---an 
impulse term and 
a similar diagram from the delta isobar--- 
are anomalously small and partly cancel. 
The bulk of the cross-section must then arise from contributions that 
are relatively unimportant in other processes. 
One is isoscalar pion rescattering for which two sets of $\chi$PT
parameters were used: 
``ste'' from a sub-threshold expansion of 
the $\pi N$ amplitude and ``cl'' from an one-loop analysis of threshold 
parameters. Others are two-pion exchange and short-range $\pi NNNN$ terms, 
which were modeled by heavier-meson exchange: 
pair diagrams with $\sigma$ and $\omega$ exchange, and 
a $\pi\rho\omega$ coupling, among other, smaller terms.    
Two potentials ---Argonne V18 and Reid93--- were used. 
Results are shown in Fig. \ref{F:vkolck:mesexfig3} 
together with IUCF and Uppsala data.
Other $\chi$PT studies of this reaction can be found in 
Ref. \cite{vkolck:pppppizero},
while Ref.\cite{vkolck:park2} presents a related analysis of the 
axial-vector current. 

The situation here is clearly unsatisfactory, and presents therefore a unique 
window to the nuclear dynamics. Work is in progress, for example, on
a similar analysis for the other, not so suppressed channels
$\rightarrow d\pi^+, \rightarrow pn\pi^+$ \cite{vkolck:carocha}.

\section*{Conclusions}

Mesonic $\chi$PT is now a mature subject, where the validity
of the approach from both phenomenological and internal consistency
standpoints has been demonstrated. $\chi$PT has also passed several tests
in systems with one nucleon, even though some issues ---such as delta isobar
effects--- remain to be fully investigated.

In nuclear physics, only the very initial steps of 
a systematic chiral expansion have been attempted
so far, despite the amount of information available.
I have tried to argue that the first results are very auspicious.
The chiral expansion has the basic ingredients of nuclear forces, as evidenced
by a quantitative fit to two-nucleon data and by the qualitative insights
into the size of few-body and isospin-violating forces. 
It provides a consistent framework for scattering on the nucleon and
on light nuclei, which in turn offers a handle on nucleon parameters
(as for pion-deuteron scattering and pion photoproduction),
successful quantitative postdictions (such as in radiative neutron-proton 
capture), and
quantitative predictions (such as in pion photoproduction). 
And best of all, it has open problems 
such as pion production in the $pp$ reaction.  
There is still a lot to be done: 
consistent potential/kernel calculations,
the above processes away from threshold,
many other processes, 
extension to $SU(3)$ and nuclear matter, to mention just a few topics.
Perhaps $\chi$PT will then fulfill the role of a long-lacking 
theory for nuclear physics based on QCD.

\vspace{1cm}
\paragraph*{Acknowledgements.}
I am grateful to my collaborators for helping making this research program
possible. This manuscript benefitted from communications with 
E. Korkmaz and T.-S. Park, and criticism by P. Bedaque.
This research was supported by the DOE grant DE-FG03-97ER41014.

\end{document}